# InGaAs and GaAsSb PIN Nanowire Arrays with Large Absorption around 1.55 µm for Room Temperature Infrared Applications

**Shrivatch Sankar[1], Punam Murkute[1], Nathan Gajowski[1], Micah Meleski[1], Md. Saiful Islam Sumon[1], John Mcelearney[2], Lawrence Qiu[2], Neha Nooman[1], Paul J. Simmonds[2], Ronald M. Reano[1], Sanjay Krishna[1], Shamsul Arafin[1]***

[1] Department of Electrical and Computer Engineering, The Ohio State University, Columbus, USA
[2] Department of Electrical and Computer Engineering, Tufts University, Medford, USA

*Author to whom any correspondence should be addressed.

E-mail: arafin.1@osu.edu

Received xxxxxx
Accepted for publication xxxxxx
Published xxxxxx

### Abstract

We report a comparative analysis between InGaAs and GaAsSb-based top-down cylindrical nanowire arrays fabricated from *p-i-n* heterostructures. Optical constants measured via spectroscopic ellipsometry were incorporated into FDTD simulations to optimize array geometries. Simulations indicated that GaAsSb nanowire arrays with a diameter of 314 nm and a period of 904 nm achieved a peak absorption of ~98% at 1550 nm, while InGaAs nanowire arrays with a 326 nm diameter and 1090 nm period reached ~76%. Measurements of fabricated arrays showed absorption of ~89% at 1628 nm (GaAsSb) and enhanced sub-bandgap absorption at 1766 nm (InGaAs) at room temperature. These results were confirmed through modelling of the as-fabricated nanowire diameters 330 nm and 411 nm, demonstrating excellent agreement between theory and experiment. In both material platforms, nano-structuring improved the absorption relative to planar films, demonstrating near-unity absorption suitable for practical single-photon detection.

## 1. Introduction

Single-photon detectors (SPDs) are crucial for advancing quantum information science (QIS), enabling a wide range of applications from secure quantum key distribution (QKD) to long-range quantum communication networks and advanced quantum sensing [1-4]. The C-band, centered around 1.55 µm, is particularly important for these technologies, as it corresponds to the minimum attenuation window of silica optical fibers [5], allowing efficient long-distance transmission of photonic qubits [6-8]. As a result, developing highly sensitive SPDs that operate in this wavelength range is a key priority for the practical deployment of quantum systems [9, 10].

Currently, superconducting nanowire single-photon detectors (SNSPDs) are considered the benchmark for high-performance single-photon counting at 1.55 µm, offering high detection efficiency and low dark counts [11-13]. Despite their exceptional detection efficiency, SNSPDs rely on bulky, power-intensive cryogenic operating at temperatures of a few kelvin, or even as low as millikelvin, which fundamentally limits their scalability and integration into practical systems [14-16]. These constraints hinder their use in distributed quantum networks, satellite-based communication, and other applications that require compact, cost-effective, and room-temperature (RT) operation.

To overcome the limitations of SNSPDs, the III-V semiconductor-based single photon nanowire array detector presents a compelling solution for RT single photon detection, with properties such as high absorption coefficients, excellent carrier mobility, reduced dark current, and broader spectral response compared to conventional Si-based SPDs [17-19]. Materials such as InGaAs and GaAsSb are particularly well-suited for engineering III-V SPDs at 1.55 μm wavelength due to their direct bandgaps and favorable electronic properties. Additionally, both InGaAs and GaAsSb can be grown lattice-matched to InP substrates, which facilitates seamless integration with established photonic platforms. While bulk III-V photodetectors require a thick absorption layer to achieve high quantum efficiency, nanowire arrays can achieve high absorption in thinner layers, leading to potential device bandwidth advantages, making them a compelling alternative.

Structuring III-V materials into nanowire arrays creates an optical metamaterial with enhanced optical properties, significantly boosting absorption efficiency within a compact volume [18, 19]. The sub-wavelength nature of the nanowire arrays allows for effective light trapping, a phenomenon where incident light is diffracted and coupled into the array. This process excites low-loss leaky guided modes within the individual nanowires, trapping the light and forcing it to propagate along the nanowire length [20, 21]. The increased optical path length within the absorbing material, combined with the resonant localization of the electromagnetic field, leads to a substantial increase in the absorption cross-section far beyond the nanowires' physical dimensions [22, 23]. Additionally, the graded effective refractive index of the nanowire array minimizes Fresnel reflection at the air-nanowire interface, acting as a broadband anti-reflection layer [19, 24, 25]. Previous research has demonstrated the potential of III-V nanowire detectors. For instance, single InP and InGaAs nanowires have been explored for photodetectors, highlighting their high sensitivity and fast response times [18, 19, 26]. Similarly, studies on InGaAs-based nanowire arrays have shown promising results in enhancing absorption at specific wavelengths [27-29]. However, a comprehensive, direct comparative study of both InGaAs and GaAsSb nanowire arrays, utilizing an identical and optimized fabrication process, has been largely absent from the literature. Such a side-by-side comparison is crucial for systematically evaluating the material-specific performance and for informing the selection of the most suitable platform for high-efficiency detectors at 1.55 μm. Additionally, a top-down fabrication approach provides excellent control over nanowire position, dimensions, and uniformity, enabling scalable integration with existing semiconductor manufacturing processes. Beyond their absorption advantages, nanowire Beyond their absorption advantages, nanowire structures also offer key device-level benefits, including strong carrier confinement, high-speed carrier extraction and lower timing jitter. By leveraging internal gain mechanisms and enhanced light-matter interaction, these arrays can achieve single-photon sensitivity even at room temperature with very high gain. Such properties make them ideal for a wide range of emerging applications, such as time-correlated photon-counting detector in quantum technologies, single-photon LiDAR for autonomous sensing, and deep-space optical communications, where extreme sensitivity and low noise are critical [19, 30-33].

This paper addresses this critical gap by presenting a comprehensive, comparative study of InGaAs and GaAsSb top-down nanowire arrays designed to achieve near-unity absorption at short-wave infrared (SWIR) wavelengths. Our fabricated nanowire arrays with diameters of 411 nm (InGaAs) and 330 nm (GaAsSb) achieve absorption levels of 85% and 89% at 1766 nm and 1628 nm, respectively. The novelty of this work lies in three key areas: (1) a systematic comparison of the absorption properties of InGaAs and GaAsSb nanowire arrays, which is crucial for selecting optimal materials for high-performance detectors; (2) the development of a robust and reproducible fabrication process for creating highly uniform cylindrical nanowire structures from these materials; and (3) the detailed correlation between FDTD simulations and experimental absorption measurements, providing a valuable feedback loop for validating the design principles and optimizing performance toward unity gain. This work not only provides a reproducible fabrication and design methodology but also offers crucial insights into the material-dependent performance of nanowire metamaterials. These findings pave the way for scalable RT photodetectors for next-generation quantum technologies.

## 2. Results and Discussion

### *2.1 Growth*

Two 1400 nm thick *p-i-n* samples of lattice-matched $In_{0.53}Ga_{0.47}As$ and $GaAs_{0.5}Sb_{0.5}$ are epitaxially grown on a semi-insulating, double-side polished InP substrate. The growth process utilizes molecular beam epitaxy (MBE) in a RIBER Compact 21 DZ system. InGaAs and GaAsSb samples are deposited on the InP substrate at a temperature of 510°C and 485°C, respectively. For the $In_{0.53}Ga_{0.47}As$ layer, the V/III flux ratio is maintained at 25, with an In/Ga flux ratio of 18 to achieve optimal material composition and crystalline quality. For $GaAs_{0.5}Sb_{0.5}$ growth, the As/Sb ratio was 4.7, and the III/V ratio was 17.4, designed to stabilize the Sb incorporation and achieve the desired stoichiometry. Both the as-grown planar structures are fabricated into cylindrical nanowire arrays via top-down etching, as shown in Figure 1(a). High-resolution X-ray diffraction (XRD) is performed along the 004-crystal plane, and the rocking curve measurements confirm a good lattice match of the InGaAs and GaAsSb layers with respect to InP, as shown in Figure 1(b). The RT photoluminescence (PL) measurements were performed on both samples to confirm optical activity with minimal defect states and high optical quality. Both bulk samples exhibit RT PL, with peaks at ~ 1690 nm and 1676 nm revealing their respective optical bandgap, as shown in Figure 1(c).

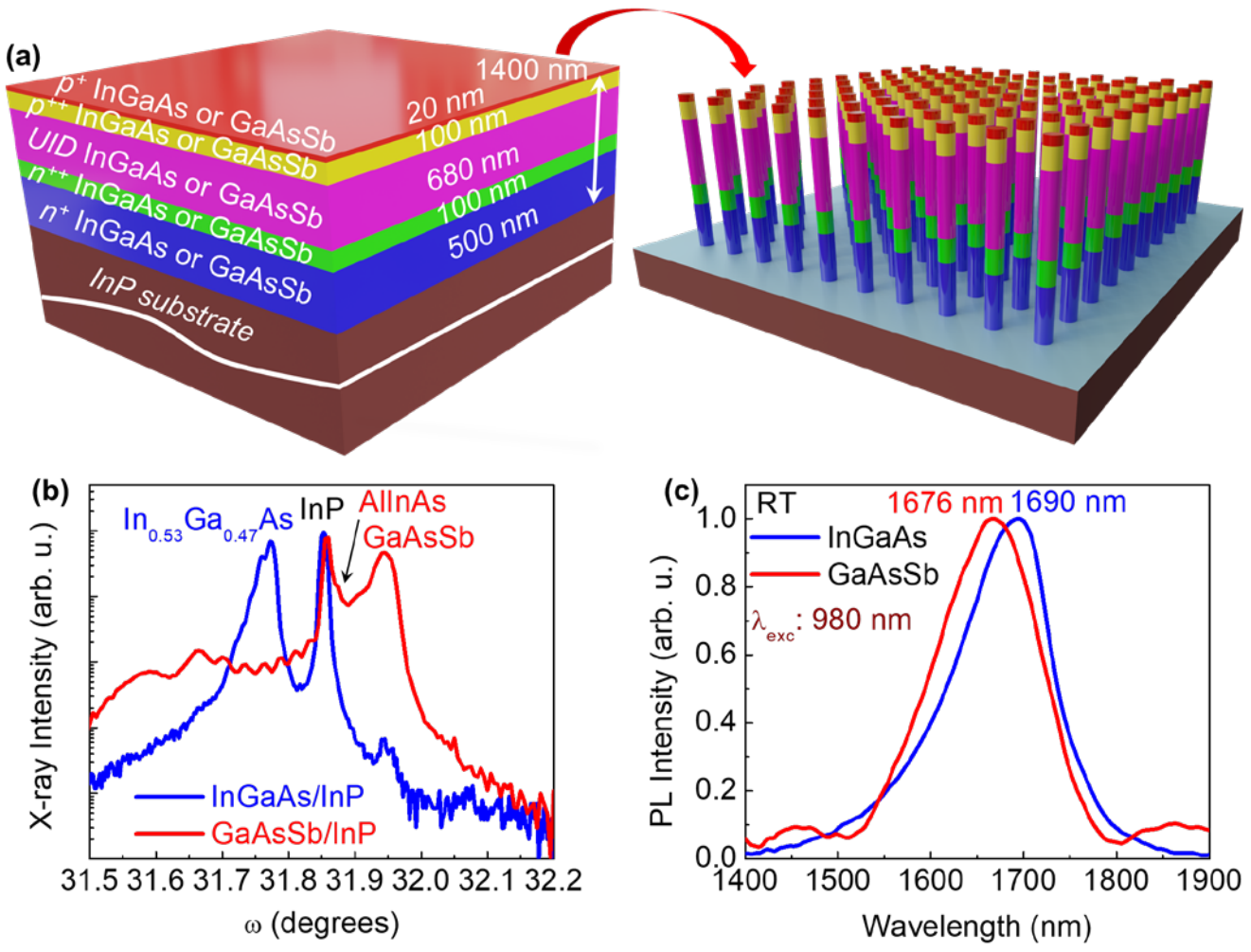

**Figure 1.** (a) Schematic of the planar InGaAs and GaAsSb *p-i-n* epitaxial structure used in this study, followed by a schematic of the top-down nanowire formation process. (b) XRD rocking curve (ω) for both the InGaAs (blue line) and GaAsSb (red line) *p-i-n* samples, and (c) photoluminescence spectra of the as-grown InGaAs (blue line) and GaAsSb (red line) *p-i-n* samples.

## *2.2 Spectroscopic Ellipsometry*

To perform electromagnetic simulations of our nanowire arrays, we require the refractive index ($n$) and extinction coefficient ($k$) of the III-V materials. These two parameters serve as the primary inputs to our photonic simulation software and form the basis for optimizing the nanowire arrays toward unity-gain absorption. The n-k data used in this simulation study for InGaAs and GaAsSb *p-i-n* and unintentionally doped (UID) structures are experimentally measured from as-grown planar samples using spectroscopic ellipsometry, as shown in Figures 2(a) and 2(b). This experimental determination is necessary due to inconsistencies in reported literature values, particularly for GaAsSb, for which refractive-index data are scarce and not consistently reported.

Spectroscopic ellipsometry (SE) measurements were performed using a J. A. Woollam Variable Angle Spectroscopic Ellipsometer (VASE), covering an energy range of 0.38-6.2 eV (200-3200 nm). Scans were taken in air at room temperature, with incident angles of 70° and 75°. For each measurement, a step size of 10 meV was used across the entire spectral range. The SE measures change in the polarization state of monochromatic light reflected from a sample, from which the material's complex refractive index ($n + ik$), as a function of wavelength, can be inferred. The incident light is decomposed into a known combination of s- and p-polarized components; because these components reflect differently depending on the refractive index of the sample, their ratio, normalized to their initial values, can be used to determine the refractive index. This ratio, ρ, is defined as [34].

$$\rho = \frac{E_{r,p}/E_{i,p}}{E_{r,p}/E_{i,s}} = \frac{r_p}{r_s} = tan(\psi)e^{i\Delta}$$

Where tan(ψ) represents the ratio of the normalized amplitudes of the reflected s- and p-polarized light, and Δ is the phase difference between them. These ellipsometric angles are then used to construct a pseudodielectric function by treating the entire sample as a single, monolithic material, using

$$\langle\varepsilon\rangle = \sin^2(\varphi) * \left[1 + \tan^2(\varphi) * \left(\frac{1-\rho}{1+\rho}\right)^2\right]$$

where φ is the angle of incidence. To extract the refractive index of a particular layer, a model is constructed, and its simulated refractive index fits the measured pseudodielectric function. Isolating the contribution of the layer of interest requires prior knowledge of the refractive indices and thicknesses of all other layers present. Accordingly, preliminary measurements of an AlInAs buffer layer were performed, while values from the J. A. Woollam library were used for the InP substrate. The unknown layer was then modeled using a series of oscillators based on common line shapes observed in the refractive indices of semiconductors [35]. Each sample discussed here also included a Bruggeman effective medium approximation (EMA) surface layer to model the effects of surface roughness and thin native oxides [36].

## *2.3 Nanowire Array Simulations*

Cylindrical nanowire arrays can function as optical metamaterials, exhibiting strong resonant behavior. The specific optical resonances that occur are primarily determined by the nanowire diameter and the array period (pitch), which define the structure's effective refractive index and govern how photons interact with the array. This tailored nanowire structure enables the excitation of leaky-mode resonances (also known as guided-mode resonances), which occur when incoming light couples into modes partially confined within the individual nanowires, such as the fundamental hybrid $HE_{11}$ guided mode. This coupling enhances light-matter interaction by trapping photons and creating strong electromagnetic fields inside and around the nanowires. A key feature of a well-designed nanowire array is its ability to suppress back-reflection, often acting as a near-perfect anti-reflection coating. This property arises from two main effects: graded refractive index, where the array creates a gradual transition in the effective refractive index from the surrounding air to the substrate material, minimizing abrupt changes that cause reflection; and resonant scattering, where leaky-mode resonances efficiently trap light within the array, ensuring it is absorbed rather than reflected and thus significantly improving absorption. By carefully tuning the nanowire diameter and period, these resonant modes can be optimized to achieve near-unity absorption over a desired wavelength range. Additionally, as these resonances are geometry-dependent, the strong field intensities they produce can significantly amplify weak absorption mechanisms, enabling the detection of light at energies below the material's intrinsic bandgap, a critical feature for applications such as sub-bandgap photodetection.

To optimize the nanowire geometries, Ansys Lumerical's 3D finite-difference time-domain (FDTD) simulation is used with periodic boundary conditions to model and analyze the nanowire array. A thorough systematic parametric sweep is performed in FDTD to tailor the nanowire diameter and period for both materials. The period of the nanowires is varied from 100 nm to 1300 nm, and the nanowire diameters range from 0 nm to 1000 nm, while the nanowire height is fixed at 1400 nm. This simulation is carried out on a free-standing nanowire array without a substrate, given the InP's transparency over 1000 nm. Nanowires with diameters greater than 400 nm exhibited higher-order HE modes, resulting in reduced absorptance. Based on the absorption results for different periodicities, the optimal period yielding near-unity absorption at a wavelength of 1550 nm is found to be 1090 nm for InGaAs and 902 nm for GaAsSb. The corresponding 2D absorption profiles are displayed in Figures 2(c) and 2(d).

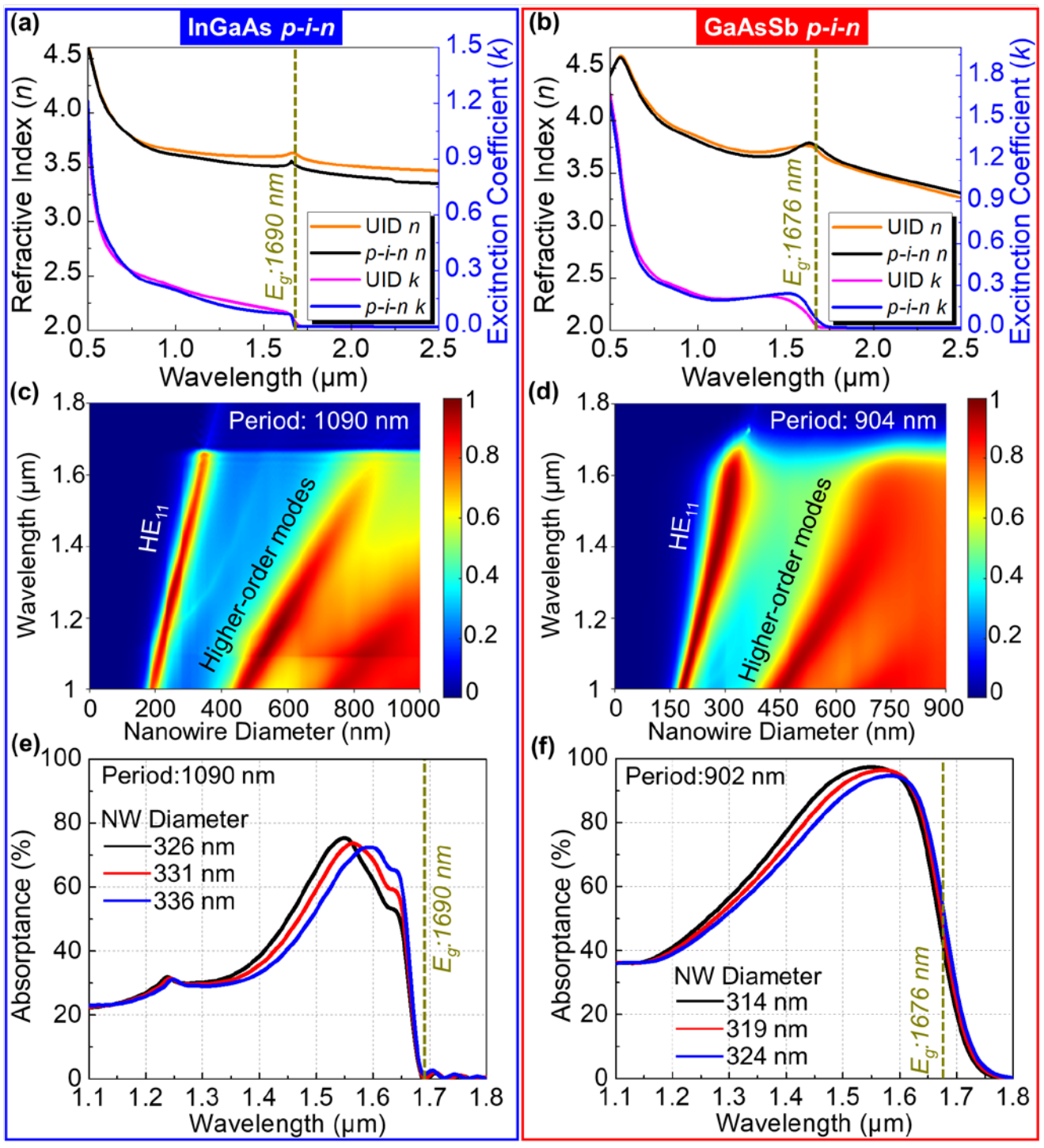

Figure 2. (a-b) Experimentally measured n and k for the (a) InGaAs and (b) GaAsSb p-i-n and UID samples, (c–d) simulated 2D absorption intensity profiles for the optimized nanowire array period of (c) InGaAs and (d) GaAsSb. The optimized array periods are similar: 1100 nm for InGaAs and 1080 nm for GaAsSb. (e–f) Simulated 2D absorption line plots for various nanowire diameters of (e) InGaAs and (f) GaAsSb, using the respective optimized nanowire array periods.

After optimizing the period for each material system, the nanowire diameter is further fine-tuned to precisely align the resonance peak with the target wavelength of 1550 nm while maximizing absorption efficiency. This fine-tuning not only allows for precise control of the resonance position but also demonstrates the wavelength tunability of the system. As shown in the 2D absorption line plots in Figures 2(e) and 2(f), increasing the nanowire diameter results in a clear redshift of the resonance wavelength. As the diameter increases, the effective optical path length and mode confinement within the nanowire array also increases, leading to a shift of the resonance toward longer wavelengths. From the simulations, we observe that for InGaAs, a nanowire diameter of 326 nm with a fixed array period of 1090 nm results in a peak resonance at 1550 nm, achieving an absorption of approximately 75.3%, as shown in Figure 3(a). Similarly, in the case of GaAsSb, a diameter of 314 nm with an array period of 904 nm yields a critical coupling resonance, with an enhanced near unity peak absorption of around 97.3%, as shown in Figure 3(b). The respective 2D absorption intensity profiles for individual nanowires are generated as shown in Figures 3(c) and 3(d). The intensity profiles are calculated from the electric-field distribution and refractive index using the relation

$$P_{abs} = -0.5\omega|E|^2 \, imag(\varepsilon)$$

following the computational approach implemented in Lumerical. It is notable that the InGaAs nanowire array exhibits significantly lower absorption at 1550 nm compared to GaAsSb, despite achieving near-unity absorption at shorter NIR wavelengths. This disparity is primarily attributed to intrinsic material properties, particularly the extinction coefficient ($k$) of InGaAs at 1550 nm (~0.07), which is approximately three times lower than that of GaAsSb (~0.24), as shown in Figures 2(a) and 2(b). This reduction in absorption for InGaAs is also evident in bulk absorption measurements, where GaAsSb (64%) continues to dominate over InGaAs (42%) at 1550 nm. These simulations confirm that GaAsSb-based nanowire arrays are the superior material for achieving near-unity absorption in the NIR wavelength range.

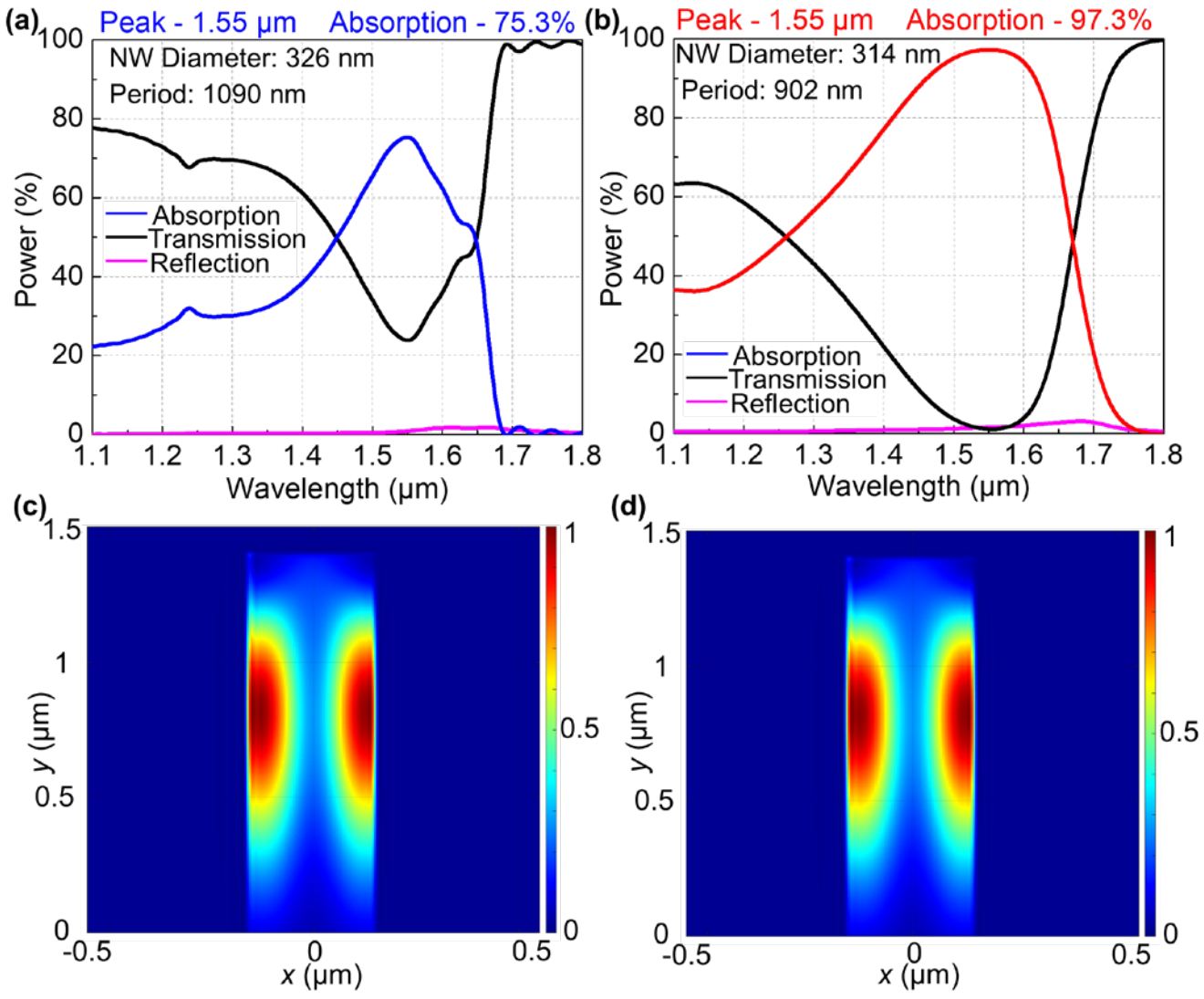

Figure 3. (a-b) 2D line plots of the optimized near-unity absorption, transmission, and reflection spectra of nanowire array geometries for: (a) InGaAs, with a nanowire diameter of 326 nm and a period of 1090

nm, achieving an absorption of 75.3%; (b) GaAsSb, with a nanowire diameter of 314 nm and a period of 904 nm, achieving an absorption of 97.3%. (c-d) Simulated 2D absorption intensity profiles for the optimized nanowire array geometries of: (c) InGaAs and (d) GaAsSb.

### *2.4 Nanowire Array Fabrication*

From the simulations, the optimized designs are used to fabricate nanowires from planar InGaAs and GaAsSb samples. The fabrication of the nanowire arrays involves a four-step process, starting with silicon nitride ($SiN_x$) hard mask deposition, followed by electron-beam lithography (EBL) patterning and development, hard-mask plasma etching, and finally III-V plasma etching for nanowire array formation. The dimensions of each nanowire array are 250 µm × 250 µm. For InGaAs nanowire formation, a $CH_4/Cl_2/H_2$-based reactive ion etching (RIE) plasma etch process with parameters including gas flow 10/15/10 sccm, chamber pressure 20 mTorr, ICP/RIE power 750/85W, duration 240 sec is used due to its high etch rate and smooth sidewall passivation. The process is carried out at a high temperature of 180 °C to suppress the formation of indium chloride ($InCl_x$) residues on the nanowire sidewalls. In contrast, high-temperature processing was initially considered unnecessary for GaAsSb due to the absence of indium. Two different etch chemistries, $BCl_3$/Ar and $Cl_2/N_2$, were explored. However, $BCl_3$/Ar yielded poor selectivity (~1:3) between the III–V material and the $SiN_x$ mask and resulted in rougher sidewalls. Consequently, $Cl_2/N_2$ plasma chemistry was pursued. While it offered moderate selectivity, it led to undesirable nanowire sidewall tapering. A detailed parametric study was conducted by varying etching parameters such as ICP/RF power, chamber pressure, and gas flow rates. Despite these efforts, vertical sidewalls could not be achieved using the $Cl_2/N_2$ recipe. Therefore, the high-temperature InGaAs etch process with pressure 15 mTorr was employed for 112 sec to achieve vertical and smooth GaAsSb nanowire sidewalls. The final fabricated InGaAs and GaAsSb nanowire arrays, demonstrating optimized etch profiles and structural fidelity, are shown in the SEM images in Figures 4(a) and 4(b). Cylindrical InGaAs nanowires with a diameter of 411 nm and an etch depth of 1381 nm are formed. An increase in nanowire diameters from 320 nm to 411 nm was observed here possibly due to re-deposition of etch products within the sidewalls during dry etching. For GaAsSb, cylindrical nanowires with a diameter of 330 nm and an etch depth of 1370 nm were fabricated. Unlike InGaAs, the deviation from the intended diameters is only ~16 nm in case of GaAsSb due to the absence of non-volatile byproducts during etching.

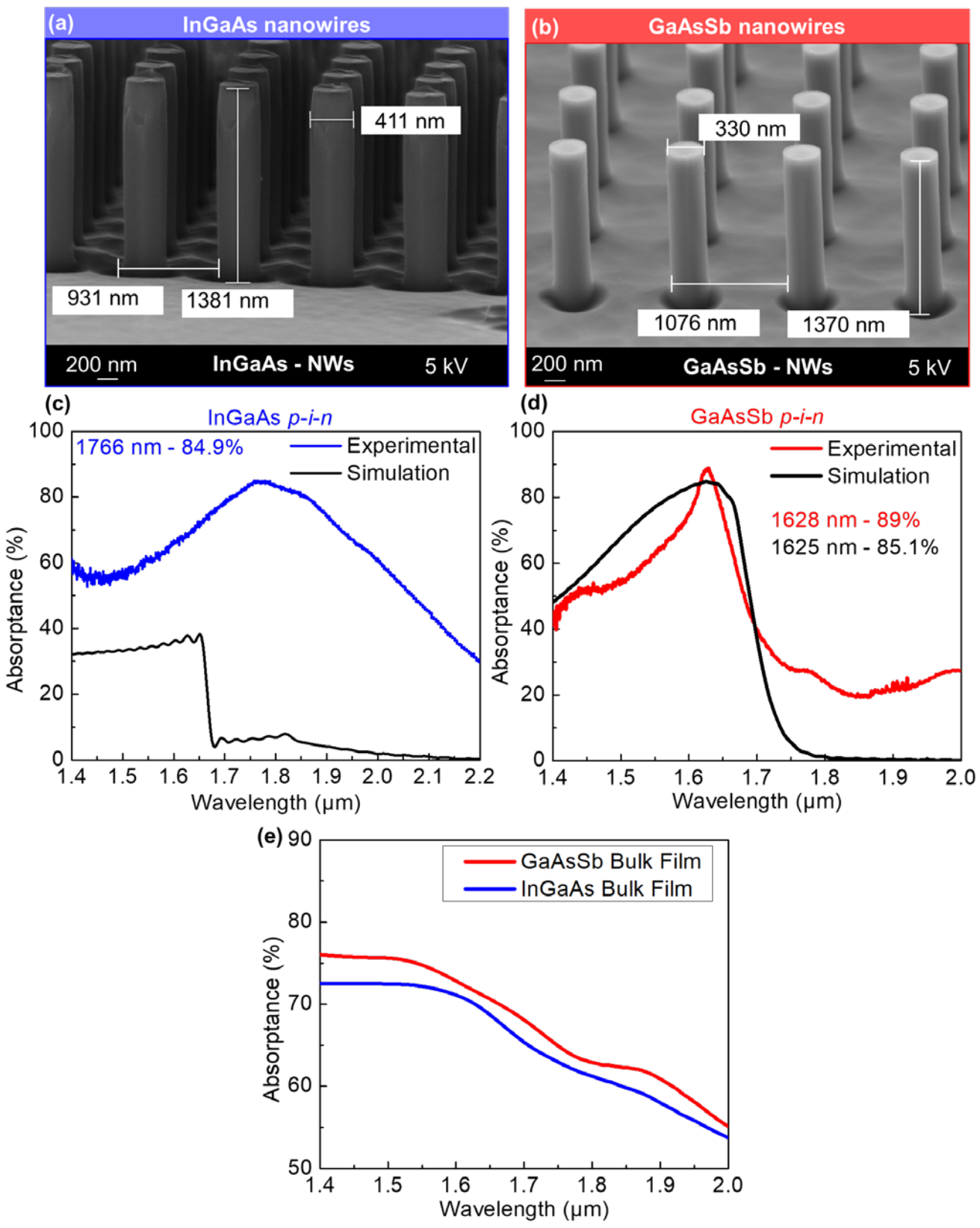

Figure 4. (a-b) Scanning electron microscopy (SEM) images of cylindrical nanowire arrays with smooth, vertical sidewalls: (a) InGaAs, (b) GaAsSb. (c-d) Percentage absorption measured in the fabricated and simulated nanowire arrays: (c) InGaAs, (d) GaAsSb, (e) Percentage absorption measured in the bulk (e) InGaAs and GaAsSb.

### *2.5 Absorption Measurement*

Transmission and reflection measurements are conducted on fabricated InGaAs and GaAsSb nanowire arrays at RT, and absorption is calculated using the relation $A = 1\text{-}T\text{-}R$. During these measurements, substrate effects are eliminated to accurately measure the nanowire array's absorption. Both nanowire types demonstrate high peak absorption efficiencies, with InGaAs nanowires reaching approximately 84.9% at 1750 nm and GaAsSb nanowires achieving about 89% at 1628 nm, as shown in Figures 4(c) and 4(d). Although the PL peak intensities for both materials are centered around 1700 nm, the maximum absorption occurs at different wavelengths, highlighting the material-dependent tunable optical responses of the nanowire arrays. Notably, both InGaAs and GaAsSb nanowires exhibit significantly higher absorptance compared to their bulk film counterparts, underscoring the critical role of nanostructuring and geometry in enhancing optical absorption.

To evaluate the consistency between theoretical and experimental performance, the measured absorption spectra are compared with simulations (Figure 4). Initial discrepancies arise from slight differences between the designed (Figure 3) and fabricated nanowire dimensions, requiring revised post-fabrication modeling based on the actual structural parameters. For the InGaAs nanowire array, the simulated absorption initially aligns with experimental trends, rising to 40% at 1650 nm. However, beyond this wavelength, the simulated profile falls abruptly and approaches zero around 1680 nm. This drops close to 1680 nm directly matches the extinction coefficient ($k$) data (Figure 2a), indicating the material's effective bandgap. While SE accurately measures $n$ and $k$ near the band edge using harmonic oscillator models, it directly assumes zero absorption in the sub-bandgap regime due to the lack of accurate theoretical models. In contrast, our experimental results (Figure 4c) show a peak absorption of 85%, clearly demonstrating that nanowire arrays can effectively extend the material's absorption range more than 100 nm beyond its intrinsic bandgap. This highlights the array's ability to efficiently trap sub-bandgap photons, offering significant tunability for broader applications. In contrast, the GaAsSb array shows near-perfect agreement between experimental and simulated data, with both exhibiting a resonance peak around 1628 nm and ~85-89% absorptance. This strong agreement confirms the high accuracy of our experimentally modeled optical constants ($n$ and $k$) as well as our EM simulation model, since the 1628 nm wavelength lies within the material's fundamental absorption band, where the SE models remain highly accurate. The FWHM of GaAsSb nanowires exhibits a narrower optical response compared to InGaAs, while simulations predicted a broader response. This spectral narrowing likely results from reduced modal overlap, which limits the number of coupled resonant modes, leading to a dominant single-mode absorption. Notably, the strong absorption peak observed well below the bandgap in the InGaAs array emphasizes the unique functionality of these nanostructures.

**Table 1.** Key results of this nanowire-array study

| Nanowire Parameters | InGaAs *p-i-n* Nanowire Array | | | GaAsSb *p-i-n* Nanowire Array | | |
|---|---|---|---|---|---|---|
| | Design | Fabricated | Redesign | Design | Fabricated | Redesign |
| Diameter | 326 nm | 411 nm | 411 nm | 312 nm | 330 nm | 330 nm |
| Array Period | 1090 nm | 931 nm | 931 nm | 902 nm | 1076 nm | 1076 nm |
| Height | 1400 nm | 1381 nm | 1381 nm | 1400 nm | 1370 nm | 1370 nm |
| Peak Wavelength | 1550 nm | 1750 nm | - | 1550 nm | 1628 nm | 1625 nm |
| Peak Absorption | **75.3%** | **84.9%** | **-** | **97.3%** | **89%** | **85.1%** |

## 3. Conclusion

This work presents the successful design, fabrication, and characterization of top-down cylindrical InGaAs and GaAsSb nanowire arrays optimized for near-unity absorption in the SWIR range. Key results are summarized in Table 1. A major outcome is the rigorous validation of our integrated simulation-fabrication process flow. The GaAsSb array exhibits excellent agreement between simulation and experiment, confirming both the accuracy of the optical modeling and the reproducibility of the nanostructure geometry. The InGaAs array achieved a similarly high absorption efficiency at 1750 nm (~85%), deep within the sub-bandgap regime. This underscores a critical advantage of the nanowire geometry's ability to effectively extend the operational bandwidth beyond the material's bulk limits. Additionally, both nanowire arrays substantially outperformed their bulk counterparts Figure 4(e), underscoring the effectiveness of nano-structuring in enhancing SWIR absorption. Importantly, for applications targeting near-unity absorption in the SWIR region, GaAsSb is a superior choice over InGaAs due to its higher intrinsic absorption. Overall, these results establish a robust and scalable approach for engineering high-performance photonic devices. By demonstrating a reproducible fabrication process validated by high-fidelity modeling, this work confirms the potential of cylindrical semiconductor nanowires for quantum photodetection and paves the way toward efficient, tunable, and scalable SWIR photodetectors that operate without cryogenic cooling.

## Data availability statement

The data that support the findings of this study are available upon reasonable request from the corresponding author.

## Acknowledgments

This work was supported by the College of Engineering, The Ohio State University (NSF) under Strategic Research Grant 134827.

## Conflict of interest

The authors declare no conflicts of interest.